\documentclass[aps,prl,showpacs,noshowkeys,amsmath,amssymb,amsfonts,reprint]{revtex4-1}
\usepackage{graphicx}
\usepackage{bm}

\usepackage[colorlinks=true,linkcolor=blue,citecolor=blue,urlcolor=blue]{hyperref}
\usepackage[all]{hypcap}
\begin{document}
\title{Fluidization and Active Thinning by Molecular Kinetics in Active Gels}
\author{David Oriola} \email{oriola@mpi-cbg.de, oriola@pks.mpg.de}
\altaffiliation{Present address: Max Planck Institute of Molecular Cell Biology and Genetics, Pfotenhauerstra{\ss}e 108, 01307  \& Max Planck Institute for the Physics of Complex Systems, N\"othnitzerstra{\ss}e 38, 01187, Dresden, Germany.}
\affiliation{Departament de F\'isica de la Mat\`eria Condensada and Universitat de Barcelona Institute of Complex Systems (UBICS), Universitat de Barcelona, Avinguda Diagonal 647, 08028 Barcelona, Catalonia, Spain}
\author{Ricard Alert}\email{ricardaz@ecm.ub.edu}
\affiliation{Departament de F\'isica de la Mat\`eria Condensada and Universitat de Barcelona Institute of Complex Systems (UBICS), Universitat de Barcelona, Avinguda Diagonal 647, 08028 Barcelona, Catalonia, Spain}
\author{Jaume Casademunt}
\affiliation{Departament de F\'isica de la Mat\`eria Condensada and Universitat de Barcelona Institute of Complex Systems (UBICS), Universitat de Barcelona, Avinguda Diagonal 647, 08028 Barcelona, Catalonia, Spain}
\date{\today}


\begin{abstract}
We derive the constitutive equations of an active polar gel from a model for the dynamics of elastic molecules that link polar elements. Molecular binding kinetics induces the fluidization of the material, giving rise to Maxwell viscoelasticity and, provided that detailed balance is broken, to the generation of active stresses. We give explicit expressions for the transport coefficients of active gels in terms of molecular properties, including nonlinear contributions on the departure from equilibrium. In particular, when activity favors linker unbinding, we predict a decrease of viscosity with activity --- active thinning --- of kinetic origin, which could explain some experimental results on the cell cortex. By bridging the molecular and hydrodynamic scales, our results could help understand the interplay between molecular perturbations and the mechanics of cells and tissues.
\end{abstract}
                              
\maketitle
Active polar gels are viscoelastic media made out of orientable constituents endowed with an internal source of energy under nonequilibrium conditions \cite{Julicher2007,*Joanny2009,Prost2015}. These materials are common in cell and tissue biology, with a prominent example being the actomyosin cortex of eukaryotic cells, which generates forces that enable cell shape changes and motility. This dynamic structure is a crosslinked network of actin polar filaments and myosin molecular motors that generates forces by transducing the chemical energy of adenosine triphosphate (ATP) hydrolysis. Other biological active gels include the mitotic spindle, bacterial suspensions and tissues.

The coarse-grained dynamics of such systems can be captured by the hydrodynamic equations of active polar gels \cite{Joanny2011,Marchetti2013,Prost2015}. Generic derivations of such equations are based on symmetry arguments \cite{Simha2002,Hatwalne2004,Marchetti2013} and/or on the formalism of linear irreversible thermodynamics \cite{Kruse2004,*Kruse2005,*Julicher2011,Marchetti2013}. These phenomenological approaches introduce a number of transport coefficients whose dependence on molecular quantities is not predicted. Such relations have been obtained in derivations of the hydrodynamic equations from microscopic models \cite{Marchetti2013} consisting of active filaments \cite{Liverpool2006a} or swimmers \cite{Baskaran2009a}, inspired by the cytoskeleton and bacterial suspensions, respectively. However, these microscopic descriptions may not be appropriate for other media such as epithelia, where cells rearrange while keeping confluence, thus allowing for tissue remodeling yet preserving mechanical integrity \cite{Guillot2013}. Not only in tissues \cite{Kametani2007,*Caicedo-Carvajal2010,*Garcia2015} but also in actomyosin gels \cite{LeGoff2001,*Humphrey2002}, in the actin cytoskeleton \cite{Wottawah2005,*Ehrlicher2015,*Ahmed2015}, and in the metaphase spindle \cite{Shimamoto2011}, flows are regulated by the binding dynamics of linker molecules. Although they crucially affect the properties and dynamics of these media, a connection between molecular kinetics and the transport coefficients of continuum theories remains elusive.

Here, we consider a collection of polar elements linked by elastic molecules, and derive the constitutive equations of an active polar gel from the nonequilibrium dynamics of the linkers. Hence, explicit expressions for the transport coefficients of active gels are given in terms of molecular parameters, including the deviation from detailed balance. In particular, our results unveil a dependence of viscosity on molecular activity, which could explain some experimental observations. This \emph{active thinning} phenomenon is different from the activity-dependent apparent viscosity of active fluids, which has a hydrodynamic origin \cite{Hatwalne2004,Giomi2010,Ramaswamy2010,Marchetti2013}. More generally, our approach provides a connection between macroscopic properties and underlying molecular processes in cells and tissues.

\begin{figure}[hbtp!]
\includegraphics[width=\columnwidth]{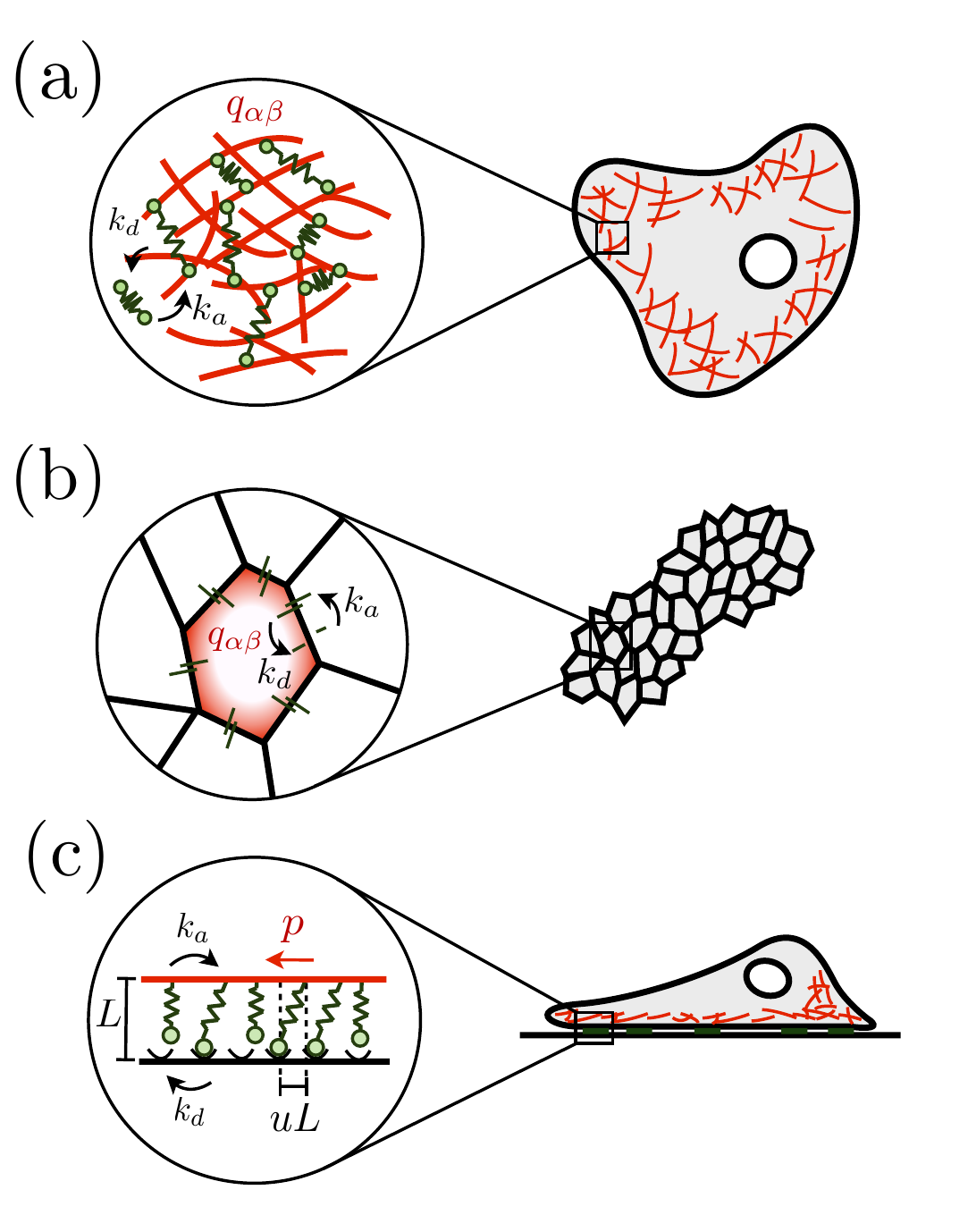}
\caption{Applications of our model to biological active gels. The elastic kinetic elements are depicted in green while the polar structures are shown in red. (a) Cell cortex: myosin motors are the active elastic kinetic elements within the actin network. (b) Tissues: cell-cell adhesion molecules, such as cadherins, are the elastic kinetic elements connecting cell cortices into a multicellular active polar gel. (c) Lamellipodium: cell adhesion molecules, such as integrins, are the elastic kinetic elements at the interface of the actomyosin gel layer.}
\label{fig1}
\end{figure}

{\it Bulk constitutive equations.} --- First, we derive the constitutive equations in the bulk of an active polar gel, e.g. in the cell cortex or in tissues (Fig. \ref{fig1}a, b). To this end, we consider a $d$-dimensional polar assembly (the actin network or the cell colony, respectively, in red) with an orientation characterized by the coarse-grained nematic order parameter field $q_{\alpha\beta} = p_\alpha p_\beta - p^2\delta_{\alpha\beta}/d $, where $p_\alpha$ is the coarse-grained polarity vector. The polar elements are crosslinked by a density $\rho$ of elastic molecules (e.g. myosin motors or cadherins in Fig. \ref{fig1}a,b, respectively, in green), so that the composite is an elastonematic material. Assuming an isotropic linear elastic response of the molecules, the free energy density of small shear deformations \footnote{The procedure and results including bulk deformations are discussed in \cite{SM}.} reads $f= \mu/2 \,u_{\alpha\beta}u_{\alpha\beta}+D\,u_{\alpha\beta}q_{\alpha\beta}+ \chi/2 \, q_{\alpha \beta} q_{\alpha \beta}$ to lowest order in $u_{\alpha \beta}$ and $q_{\alpha \beta}$, being $u_{\alpha\beta}$ the (symmetric and traceless) strain tensor, $\mu$ the shear elastic modulus, $D$ the elastonematic coefficient, and $\chi$ the inverse nematic susceptibility \cite{Lubensky2002}. Thermodynamic stability, namely convexity of the free energy, imposes $\mu\chi>D^2$.

Assuming spatial uniformity, we define $n\left(\mathbf{u},\mathbf{q},t\right)d\mathbf{u} \, d\mathbf{q}$ as the fraction of bound molecules with strain $\left[\mathbf{u},\mathbf{u}+d\mathbf{u}\right]$ and nematic order $\left[\mathbf{q},\mathbf{q}+d\mathbf{q}\right]$ at time $t$. Then, the stochastic linker dynamics is captured by the following equation for the distribution of bound linkers \cite{Tanaka1992b}
\begin{equation} \label{eq master3d}
\frac{\partial n}{\partial t} + v_{\alpha\beta} \frac{\partial n}{\partial u_{\alpha\beta}} + \dot{Q}_{\alpha\beta} \frac{\partial n}{\partial q_{\alpha\beta}} = (1-\phi_b)k_a-nk_d.
\end{equation}
Here, $v_{\alpha\beta}=\left\langle\dot{u}_{\alpha \beta}\right\rangle$ and $Q_{\alpha \beta}=\left\langle q_{\alpha \beta}\right\rangle$ are the strain rate and order parameter tensors, respectively. Brackets denote ensemble averages, so that $v_{\alpha\beta}$ and $Q_{\alpha\beta}$ are the hydrodynamic variables. We assume rigid polar elements (actin fibers or cell cortices in Fig. \ref{fig1}a, b, respectively) that do not deform \cite{Julicher1997}, so that all the linkers shear and reorient at the same rate, consistently with $v_{\alpha\beta}$ and $\dot{Q}_{\alpha\beta}$ being spatially uniform. In turn, $\phi_b \equiv \int_{\mathbb{R}^m} n\,d\mathbf{u} \, d\mathbf{q}$ is the total fraction of bound molecules, with $m=d(d+1)-2$ being the total number of independent components of the strain and order parameter tensors. Finally, $k_a$ and $k_d$ are the attachment and detachment rates of the linker molecules, respectively.

In active systems, detailed balance is locally broken. This can be generically expressed as \cite{Julicher1997}
\begin{equation} \label{eq rates}
\frac{k_a}{k_d}=e^{-\beta\varepsilon}-\Omega,
\end{equation}
with $\beta\equiv\left(k_BT\right)^{-1}$. Here, $\varepsilon=\varepsilon_0+f/\rho$ is the free energy difference between the bound and unbound states per molecule, including its chemical part $\varepsilon_0$. In turn, $\Omega$ characterizes the departure from detailed balance, hereafter referred to as `activity'. It is an a priori unknown function of the parameters, scalar combinations of $u_{\alpha\beta}$ and $q_{\alpha\beta}$, and the chemical potential difference $\Delta\mu$ of ATP hydrolysis, with $\Omega \propto \Delta\mu$ close to equilibrium.

At this point, for each particular system, it is necessary to introduce the appropriate force dependence of the molecular unbinding rate $k_d$. For simplicity, and to obtain explicit expressions of the transport coefficients, we now choose a force-independent unbinding rate. This corresponds to assuming the barrier of the binding energy landscape of the molecules to be very close to the bound state \cite{Walcott2010}. Under this assumption, the stationary fraction of bound linkers $\phi_b$ is obtained by introducing Eq. \ref{eq rates} in Eq. \ref{eq master3d} and integrating over $\mathbf{u}$ and $\mathbf{q}$:
\begin{equation} \label{eq fraction}
\phi_b=\frac{\alpha-\Omega_0}{1+\alpha-\Omega_0},
\end{equation}
where $\alpha \equiv \int_{\mathbb{R}^m}e^{-\beta\varepsilon} d\mathbf{u} \, d\mathbf{q}=\left(\frac{2\pi\rho}{\beta\sqrt{\mu\chi-D^2}}\right)^{m/2} e^{-\beta\varepsilon_0}$ and $\Omega_0\equiv\int_{\mathbb{R}^m}\Omega \, d\mathbf{u} \, d\mathbf{q}$ respectively characterize the equilibrium and active parts of the molecular kinetics, with $\Omega_0< \alpha$.

The stress $\sigma_{\alpha \beta}$ of the composite network \cite{Tanaka1992b} and its nematic field $H_{\alpha \beta}$ can be defined as
\begin{subequations} \label{eq definitions}
\begin{eqnarray} 
\sigma_{\alpha\beta}&=&\int_{\mathbb{R}^m} n \,  \sigma^{\rm el}_{\alpha \beta} \, d\mathbf{u} \, d\mathbf{q}, \label{eq stress}\\
H_{\alpha\beta}&=&\int_{\mathbb{R}^m} n \,  h_{\alpha \beta} \, d\mathbf{u} \, d\mathbf{q}, \label{eq Hfield}
\end{eqnarray}
\end{subequations}
where $\sigma^{\text{el}}_{\alpha\beta}=\partial f/\partial u_{\alpha\beta}$ is the elastic stress sustained by the linkers and $h_{\alpha\beta}=\partial f/\partial q_{\alpha\beta}$ is the coarse-grained nematic field \footnote{The minus sign in these definitions is removed because we refer the quantities to the medium, not to the linkers.}. Then, computing the time derivative of Eqs. \ref{eq definitions} and using Eqs. \ref{eq master3d}-\ref{eq fraction}, we obtain the constitutive equations of the active polar gel (see details in \cite{SM}):
\begin{subequations} \label{eq bulk constitutive}
\begin{eqnarray}
\left(1+\tau \frac{d}{dt}\right) \sigma_{\alpha \beta} &=& 2\eta \, v_{\alpha \beta}- \nu \, \dot{Q}_{\alpha \beta}- \zeta\, Q_{\alpha \beta}, \label{eq bulk stress} \\
\left(1+\tau \frac{d}{dt}\right) H_{\alpha \beta} &=&  \gamma \, \dot{Q}_{\alpha \beta} + \nu \, v_{\alpha \beta} - \omega \, Q_{\alpha \beta}, \label{eq bulk Hfield}
\end{eqnarray}
\end{subequations}
where $\tau=k_d^{-1}$ is the viscoelastic relaxation time. The viscoelastic behavior stems from the linker kinetics, which fluidizes the initially elastic network, leading to a viscous response at times longer than $\tau$ \cite{Tanaka1992b}, with shear viscosity $\eta$ and rotational viscosity $\gamma$. In addition, Eqs. \ref{eq bulk constitutive} feature flow alignment terms coupling flow to orientation by the coefficient $\nu$. Finally, the equations have terms corresponding to an active stress and an active alignment, with coefficients $\zeta$ and $\omega$, respectively. These coefficients are obtained in terms of the mechanical and kinetic molecular parameters:
\begin{equation} \label{eq coefficients}
\begin{split}
\eta=\mu\phi_b/(2k_d),\quad \gamma= \chi \phi_b/k_d,\quad \nu=-D\phi_b/k_d,\\
\zeta=(D\, \Omega_q + \mu\,  \Omega_u) \left(1-\phi_b\right),\qquad\quad\\
\omega=(D \, \Omega_u+\chi \,\Omega_q) \left(1-\phi_b\right),\qquad\quad
\end{split}
\end{equation}
with $\phi_b(\alpha,\Omega_0)$ given by Eq. \ref{eq fraction}. In turn, $\Omega_u$ and $\Omega_q$ are scalars defined by
\begin{subequations} \label{eq Omegas}
\begin{eqnarray}
\int_{\mathbb{R}^m} \Omega \, u_{\alpha \beta} \, d\mathbf{u} \, d\mathbf{q} &\equiv& \Omega_{u} Q_{\alpha \beta}, \label{eq_omegau}\\
\int_{\mathbb{R}^m} \Omega \, q_{\alpha \beta} \, d\mathbf{u} \, d\mathbf{q}  &\equiv& \Omega_{q} Q_{\alpha \beta}, \label{eq_omegau}
\end{eqnarray}
\end{subequations}
where the integrals must be proportional to the only symmetry-breaking tensor available, $Q_{\alpha\beta}$. Physically, $\Omega_u$ and $\Omega_q$ correspond to an \emph{active strain} and an \emph{active orientation} induced by the departure from detailed balance, which are ultimately responsible for the shear active stress and alignment, respectively.

Eqs. \ref{eq bulk constitutive} are the constitutive equations of an active polar gel \cite{Salbreux2009}. Here, the passive transport coefficients $\eta, \nu, \gamma$ respectively emerge from the mechanical parameters in the free energy, $\mu,D,\chi$, via the fluidization induced by linker kinetics. In turn, the active coefficients $\zeta,\omega$ are constructed by coupling scalars derived from $\Omega$ to the mechanical parameters. This clearly denotes that the generation of shear active forces requires breaking rotational invariance ($Q_{\alpha\beta}\neq 0$) and detailed balance (nonzero $\Omega_u$ and/or $\Omega_q$), which is a fundamental feature of active gels \cite{Prost2015}. Finally, as in the fluidization of tissues by cell proliferation \cite{Ranft2010}, the Maxwell operator $\left(1+\tau\, d/dt\right)$ affects $\sigma_{\alpha \beta}$ and $H_{\alpha \beta}$ but not the nematic terms in Eqs. \ref{eq bulk constitutive}, differing from the form often adopted for active gels \cite{Kruse2004,*Kruse2005,*Julicher2011}.

{\it Active thinning by molecular kinetics.} --- Eq. \ref{eq coefficients} unveils the dependence of the transport coefficients on activity at the molecular level, characterized by $\Omega_0$, $\Omega_u$, and $\Omega_q$, which can be experimentally modified by tuning the ATP concentration. In the Onsager approach to the equations of active gels, such dependencies are absent at the linear level and could only arise from nonlinear flux-force couplings \cite{Prost2015}. In our derivation, in contrast, while the constitutive equations are still linear due to having restricted the free energy to lowest order, the coefficients include contributions of all orders in the activity.

\begin{figure}[bt]
\includegraphics[width=\columnwidth]{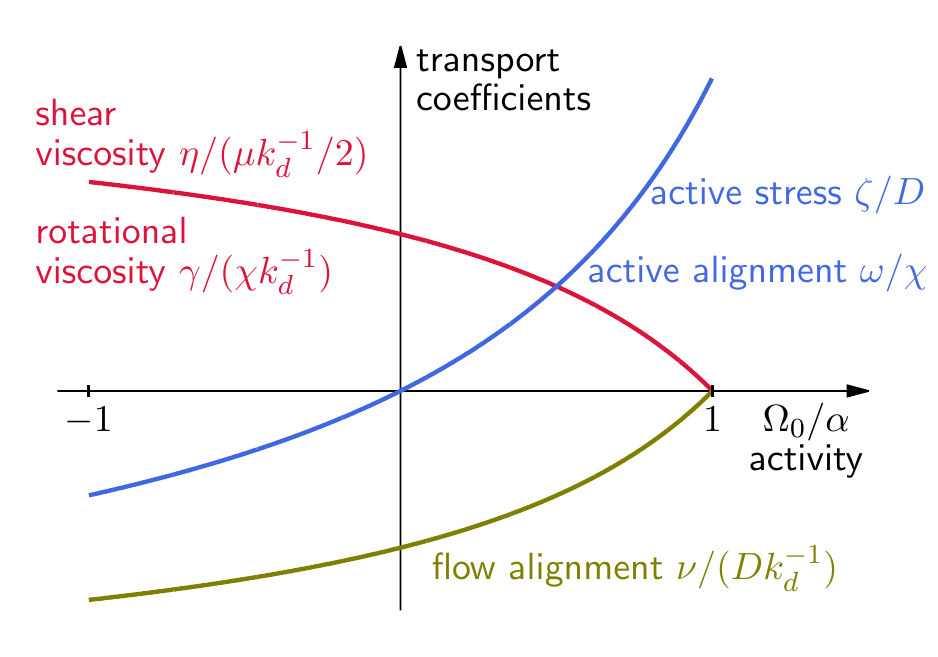}
\caption{Transport coefficients of Eqs. \ref{eq bulk constitutive}-\ref{eq coefficients} as a function of the activity $\Omega_0$, tuned by the ATP concentration, for $\Omega_u=0$ and $\Omega_q=\Omega_0$. For $\Omega_0>0$ (see text), the viscosity $\eta=\mu\phi_b/(2k_d)$ decreases with activity (\emph{active thinning}) due to the reduced fraction of bound molecules $\phi_b=\left(\alpha-\Omega_0\right)/\left(1+\alpha-\Omega_0\right)$.}
\label{fig2}
\vskip-0.2cm
\end{figure}

Fig. \ref{fig2} shows the dependence of the transport coefficients on the departure from equilibrium, $\Omega_0$, for the simple case $\Omega_u=0$ and $\Omega_q=\Omega_0$ (i.e. neglecting fluctuations of $q_{\alpha \beta}$). In general, the sign of $\Omega_0$ is not determined. For instance, for myosin, ATP binding directly causes its dissociation from actin filaments \cite{Howard2001}, suggesting that $\Omega_0>0$. For adhesion molecules such as integrins \cite{Fournier2010a} or cadherins \cite{Borghi2012}, the same behaviour may stem from the fact that activity (ATP consumption) generates cortical contractile forces that pull on them, hence favouring their detachment. However, more complex responses such as catch-bond behaviour \cite{Evans2007} might yield $\Omega_0<0$.

For $\Omega_0>0$, the viscosity decreases with activity, as shown in Fig. \ref{fig2}, which we call \emph{active thinning}. This predicted modification of viscosity with activity has a kinetic origin. Thus, it must be distinguished from the activity-dependent effective viscosity of active nematic fluids, $\eta_{\text{eff}}=\eta-\zeta \tau_q \nu/2$, with $\tau_q$ the orientational relaxation time, which is a hydrodynamic effect due to flow alignment \cite{Hatwalne2004,Giomi2010,Ramaswamy2010,Marchetti2013}. Whereas the effective viscosity depends on (the sign of) other coefficients, such as the active stress and the flow alignment, our nonequilibrium kinetic correction to viscosity is intrinsic and does not.

Consequently, activity modifies the viscosity of active gels through two different mechanisms: one based on molecular kinetics and one on flow alignment. The latter was associated to the reduction/increase of the apparent viscosity measured in active extensile/contractile suspensions of microswimmers \cite{Marchetti2013,Sokolov2009,*Rafai2010}. However, in some biological active gels, to which our linker-based model applies, the opposite effect was observed. For instance, myosin activity was shown to fluidize and soften actin gels \cite{LeGoff2001,*Humphrey2002} or even cells in suspension (lacking adhesions and stress fibers) \cite{Chan2015} and the cell cortex in mitosis \cite{Fischer-Friedrich2016}, decreasing both their stiffness and viscosity. Since actomyosin gels, and hence the cortex, are contractile ($\zeta<0$), the flow alignment effect would render an increased effective viscosity $\eta_{\text{eff}}$, which seems inconsistent with the measurements. Hence, we propose that the measured active softening could be partially due to the predicted kinetic effect (red line in Fig. \ref{fig2}), which is independent of the contractile/extensile nature of the system. ATP binding to myosin would promote its dissociation from actin and thus decrease the viscosity. Combined with increased active stress (blue line in Fig. \ref{fig2}), this would allow a network remodeling resulting in the observed fluidization.

Similar considerations might hold for suspensions of nucleic acids and proteins. Indeed, rheological measurements \cite{Hameed2012a} and the observation of collective flows \cite{Zidovska2013} suggest that chromatin behaves as a gel with active polar processes associated to chromatin remodeling enzymes \cite{Bruinsma2014}. In this line, ATP was shown to lower the apparent viscosity of nucleoli \cite{Brangwynne2011}, consistently with our prediction. Similarly, the metaphase spindle behaves as an active polar fluid \cite{Brugues2014a}, with an increased viscosity when the ATP hydrolysis rate is reduced \cite{Shimamoto2011}, also in line with our result.

{\it Interfacial constitutive equations.} --- Finally, we derive the constitutive equations at the boundary of an active polar gel, such as to account for traction forces exerted by lamellipodia on substrates via focal adhesions (Fig. \ref{fig1}c). With this purpose, we consider a polar surface (lamellipodium, red) coated with a density $\rho$ of elastic molecules (e.g. integrins, green) that transiently bind to an apolar surface (substrate, black). Now, taking the $\hat{z}$ axis perpendicular to the surface, the strain is effectively a vector $u_\alpha\equiv u_{\alpha z}$ that can directly couple to the polarity $p_\alpha$. Hence, the free energy density reads $f=\mu/2 \,u_{\alpha}u_{\alpha}+D\,u_{\alpha}p_\alpha+\chi/2 \, p_{\alpha} p_{\alpha}$, where $\mu$ is the shear elastic modulus, $D$ is the elastopolar coefficient, and $\chi$ is the inverse orientational susceptibility.

Parallel to the bulk case, the force $F_\alpha$ exerted by the bound molecules on the substrate \cite{Schwarz2013} and the average molecular field $H_\alpha$ are defined as
\begin{subequations} \label{eq interfacial definitions}
\begin{eqnarray} 
F_\alpha&=&\int_{\mathbb{R}^{k}}  n \, F^{\text{el}}_\alpha\, d\mathbf{u} \, d\mathbf{p}, \label{eq force} \\
H_\alpha&=&\int_{\mathbb{R}^{k}}  n \, h_\alpha\, d\mathbf{u} \, d\mathbf{p}, \label{eq molecular field}
\end{eqnarray}
\end{subequations}
with $k=2(d-1)$, $F^{\text{el}}_\alpha=\rho^{-1}\,\partial f/\partial u_\alpha$ being the elastic force sustained by the linkers, and $h_\alpha=\rho^{-1}\,\partial f/\partial p_\alpha$ the molecular field. Then, we find the constitutive equations at the interface of an active polar gel (see details in \cite{SM}):
\vskip-0.5cm
\begin{subequations} \label{eq interfacial constitutive equations}
\begin{eqnarray}
\left(1+\tau\frac{d}{dt}\right) F_\alpha&=&\xi_i v_\alpha - \nu_i \dot{P}_{\alpha} - \zeta_i P_\alpha, \label{eq force interfacial} \\
\left(1+\tau\frac{d}{dt}\right) H_\alpha&=&\gamma_i \dot{P}_{\alpha} + \nu_i v_{\alpha} - \omega_i P_{\alpha}, \label{eq molecular field interfacial}
\end{eqnarray}
\end{subequations}
where $P_{\alpha} = \langle p_{\alpha} \rangle$, and $v_\alpha=\langle \dot{u}_{\alpha} \rangle L$ is the gel-substrate relative velocity, with $L$ the gel-substrate distance (Fig. \ref{fig1}c). As for the bulk case, molecular kinetics entails the fluidization of the ensemble of elastic linkers, thereby leading to friction with coefficient $\xi_i=\mu\phi_b/(2k_d\rho L)$ and to an interfacial rotational viscosity $\gamma_i= \chi \phi_b/(k_d\rho)$. In turn, $\nu_i=-D\phi_b/\left(k_d\rho\right)$ is the interfacial flow alignment coefficient, and $\zeta_i=(D \, \Omega_p +\mu \, \Omega_u) \left(1-\phi_b\right)/\rho$, $\omega_i=(D\,\Omega_u+ \chi\, \Omega_p)(1-\phi_b)/ \rho$ are the interfacial active force and active alignment coefficients, respectively, with $\int_{\mathbb{R}^{k}}\Omega\,u_{\alpha} \, d\mathbf{u} \, d\mathbf{p}\equiv \Omega_u p_{\alpha}$ and $\int_{\mathbb{R}^{k}} \Omega\, p_{\alpha} \, d\mathbf{u} \, d\mathbf{p}\equiv \Omega_p p_{\alpha}$.

Eqs. \ref{eq interfacial constitutive equations} correspond to the constitutive equations at the interface of an active polar fluid (Eqs. 20-22 in \cite{Julicher2009}, omitting chemical potential gradients), thus giving their coefficients in terms of molecular parameters. A key point is that the interfacial active force $\zeta_i P_\alpha$ is polar whereas the bulk active stress $\zeta Q_{\alpha\beta}$ features apolar symmetry.

{\it Discussion.} --- We have derived the constitutive equations for the active polar gel that emerges from the nonequilibrium dynamics of a single species of elastic molecules that link polar elements. This minimal bottom-up approach is inspired by biological materials such as lamellipodia, the cell cortex, or tissues (Fig. \ref{fig1}). Assuming a constant unbinding rate of the linker molecules yields simple explicit expressions of the transport coefficients in terms of molecular parameters. In particular, the coefficients include nonlinear dependencies on activity, by means of three unknowns ($\Omega_0,\Omega_u,\Omega_q$) that characterize the departure from detailed balance. For general linker kinetics $k_d$, the approach is still valid but explicit expressions may not be obtained. Although spatial uniformity is assumed, the ensuing constitutive equations and transport coefficients can be used in the hydrodynamic limit, i.e. including small gradients.

Whereas the mechanical response of active solids had been derived from microscopic models \cite{Liverpool2009,*Banerjee2011a,*Hawkins2014}, the viscoelastic relaxation of active fluids remained only included at the hydrodynamic level \cite{Callan-Jones2011,*Hemingway2015}, thus unrelated to underlying molecular processes. In our derivation, the binding kinetics of linker molecules fluidizes the material, giving rise to a viscoelastic fluid response typically postulated in active gel theory. In general, other fluidization mechanisms may be at play, such as actin depolymerization in the cortex \cite{Chan2015}, cell division and apoptosis \cite{Ranft2010}, or topological transitions and cell shape changes \cite{Etournay2015} in tissues. We expect the fluidization mechanism associated to molecular kinetics to be generic in cells and tissues, and to combine with others in the corresponding time scales.

Building on previous works on transiently cross-linked networks \cite{Tanaka1992b,Broedersz2014}, our model accounts for orientational degrees of freedom of the gel, and explicitly includes a nonequilibrium contribution to the binding kinetics of the linkers. Active stresses and torques naturally emerge from this contribution, which also modifies the passive transport coefficients of the system. Finally, bulk and interfacial active forces are shown to exhibit different symmetries, yet depend on common parameters. Thus, in tissues, our unified treatment of intercellular (bulk) and traction (interfacial) forces may help understand their interdependence \cite{Maruthamuthu2011}. Hence, our results could shed light on active-gel models of epithelial dynamics.

More generally, our work contributes to bridging the gap between the hydrodynamics of active gels and the underlying molecular dynamics. Often, whereas macroscopic quantities such as stress and shear are measured, molecular concentrations and kinetic parameters are under experimental control \cite{Bazellieres2015}. Therefore, our results may help interpret the effects of molecular perturbations on the mechanical properties of biological active gels, from subcellular structures such as the actomyosin cortex or the mitotic spindle to multicellular tissues. In this line, we have unveiled a dependence of viscosity on ATP consumption that could explain some experimental findings.

\begin{acknowledgements}
We thank M. Rao for insightful comments and for pointing out Ref. \cite{Tanaka1992b} to us. We acknowledge C. Blanch-Mercader, P. Sens, R. Voi\-tu\-riez, J. Bru\-gu\'es, and J.-F. Jo\-an\-ny for discussions. We also acknowledge financial support from MINECO under project FIS2013-41144-P, and Generalitat de Catalunya under project 2014 SGR 878. D.O. acknowledges support from an FPU grant from the Spanish Government, and R.A. from Fundaci\'o ``La Caixa".

D.O. and R.A. contributed equally to this work.
\end{acknowledgements}

\bibliography{Main_Text}

\end{document}